\documentclass[referee]{raa_rb}
\usepackage{graphicx,times}
\usepackage{natbib}
\usepackage{amssymb,amsmath}
\usepackage{enumerate}

\bibpunct{(}{)}{;}{a}{}{,}

\begin{document}

   \title{Spatial Distributions of Sunspot Oscillation Modes at Different Temperatures$^*$
   \footnotetext{$*$ Supported by the National Natural Science Foundation of China.}
\footnotetext{\textdagger \ Corresponding author}
}

 \volnopage{ {\bf 20XX} Vol.\ {\bf X} No. {\bf XX}, 000--000}
   \setcounter{page}{1}

    \author{Zhengkai WANG\inst{1}, Song FENG\inst{1,2,}{\textsuperscript{\textdagger}}, Linhua DENG\inst{2,3,4}, Yao MENG\inst{1}
   }

   \institute{Faculty of Information Engineering and Automation, Kunming University of Science and Technology, Kunming 650500, China; \it feng.song@kmust.edu.cn \\
       \and
             Yunnan Observatories, Chinese Academy of Sciences, Kunming 650216, China\\
             \and
             College of Science, China Three Gorges University, Yichang 443002, China\\
             \and
             Chongqing University of Arts and Sciences, Chongqing 402160, China\\
\vs \no
   {\small Received 20XX Month Day; accepted 20XX Month Day}
}

\abstract{
Three- and five-minute oscillations of sunspots have different spatial distributions in the solar atmospheric layers.
The spatial distributions are crucial to reveal the physical origin of sunspot oscillations and to investigate their propagation.
In this study, six sunspots observed by Solar Dynamics Observatory/Atmospheric Imaging Assembly were used to obtain the spatial distributions of three- and five-minute oscillations.
The fast Fourier transform method is applied to represent the power spectra of oscillation modes.
We find that, from the temperature minimum to the lower corona, the powers of the five-minute oscillation exhibit a circle-shape distribution around its umbra, and the shapes gradually expand with temperature increase.
However, the circle-shape is disappeared and the powers of the oscillations appear to be very disordered in the higher corona.
This indicates that the five-minute oscillation can be suppressed in the high-temperature region.
For the three-minute oscillations, from the temperature minimum to the high corona, their powers mostly distribute within an umbra, and part of them locate at the coronal fan loop structures.
Moreover, those relative higher powers are mostly concentrated in the position of coronal loop footpoints.
}

   \authorrunning{WANG Z. et al. }            
   \titlerunning{Spatial Distributions of Sunspot Oscillation Modes at Different Temperatures}  
   \maketitle

%
\section{Introduction}           
\label{sect:intro}

Sunspot oscillations are crucial to explore structures of solar magnetic fields \citep{Bogdan+Judge+2006,DeMoortel+Nakariakov+2012}.
Two oscillation modes (one is a three-minute period, and the other five-minute) were initially found in the photosphere and chromosphere, respectively \citep{Beckers+Schultz+1972,Bhatnagar+etal+1972,Bhatnagar+Tanaka+1972,Giovanelli+1972}.
Since the early 1970s, many authors \citep[e.g.,][]{Lites+1992,Solanki+2003,Bogdan+Judge+2006,Jess+etal+2015} further studied the two modes and revealed that the oscillations exhibit different spatial distributions in different atmospheric layers \citep{Balthasar+1990,Bogdan+Judge+2006,Kobanov+etal+2011}.
The spatial distributions  are believed to be important for investigating the physical origin  of the oscillation modes and their propagation in the solar atmosphere \citep{O'Shea+etal+2002,RouppevanderVoort+etal+2003}.

From the temperature minimum to the upper coronal layer, the three-minute oscillation mostly exists in an umbra \citep{O'Shea+etal+2002}.
\cite{Nagashima+etal+2007} further stated that the three-minute oscillation can be suppressed in the photosphere and enhanced in the chromosphere.
High spatial and temporal resolution observations provided a chance to reveal spatial distributions of sunspot oscillations in different atmospheric layers.
Using data observed by Sayan Solar Observatory and Solar Dynamics Observatory/Atmospheric Imaging Assembly \citep[SDO/AIA;][]{Lemen+etal+2012,Pesnell+etal+2012}, \cite{Kobanov+etal+2013} found that the powers of the three-minute oscillation mainly concentrate within an umbra  in the temperature minimum and chromosphere.
Although the oscillation expands significantly in the transition region and lower corona,  it is still limited within the umbra.
In contrast,  \cite{Reznikova+Shibasaki+2012} stated that the three-minute oscillation leaks clearly along the coronal fan-structures in the lower corona.

Although the findings of the spatial distributions of three-minute oscillations are controversial, the main standpoints about five-minute oscillations are roughly similar to each other \citep{Lites+etal+1982,Balthasar+1990,Marco+etal+1996,Sych+Nakariakov+2008,Reznikova+Shibasaki+2012,Kobanov+etal+2013,Yuan+2015}.
The powers of  five-minute oscillations mostly distribute the umbra-penumbra boundary.
Moreover, the distribution of the powers exhibits a nearly circle-shape structure.
The structure continuously expands from the chromosphere to the transition region, but part disappears in the lower corona.

The results imply that the spatial distributions of three- and five-minute oscillations are closely related to the solar atmospheric temperature \citep{Fludra+2001,Reznikova+Shibasaki+2012}.
So, we randomly selected six sunspots observed by SDO/AIA to reveal the statistical features of sunspot oscillations at different temperatures.
The selected observations cover different temperatures from the temperature minimum to the higher corona.

The paper is as follows.  We first describe the observations and data reduction in Section \ref{Obs}, and then the oscillation modes of sunspots and their spatial distributions are given in Sections \ref{sunspotosc} and \ref{distribution}, respectively.
Finally, our conclusion is presented in Section \ref{conclusion}.

\section{Observations and Data Reduction}
\label{Obs}
All six sunspots were observed with SDO/AIA for one hour.
Table~\ref{Tab1} shows the details of these sunspots.
We chose the observations of five channels, i.e., AIA 1700 \AA\ (located in the temperature minimum), 304 \AA\ (the transition region), 171 \AA\ (the lower corona), 211 \AA\ and 131 \AA\ (the higher corona).
The characteristic temperatures of the five channels are: 0.6 $\times$ 10$^{4}$, 5 $\times$ 10$^{4}$, 100 $\times$ 10$^{4}$, 200 $\times$ 10$^{4}$, and 1000 $\times$ 10$^{4}$ K, respectively.
The image series have cadences of 12 seconds for all, except for the AIA 1700 \AA\ channel, its cadence is 24 seconds.
We used the routine $aia\_prep.pro$ to process level 1.0 data for obtaining level 1.5.
Furthermore, all the data were truncated to a region with 48\arcsec $\times$ 48\arcsec, which contains an entire sunspot.

\begin{table}
\begin{center}
\caption[]{Data series observed by SDO/AIA in our analysis}
\label{Tab1}

 \begin{tabular}{cccccc}
\hline\noalign{\smallskip}
 NOAA & Date        & Location(arcsec)    & T$_{start}$(UTC) & T$_{end}$(UTC)  \\
  \hline\noalign{\smallskip}
11176 & 27 Mar. 2011 & [1, -136]   & 16:00            & 17:00           \\
11433 & 16 Mar. 2012 & [-5, 285]   & 18:00            & 19:00           \\
11479 & 17 May 2012 & [-7, 273]   & 00:00            & 01:00           \\
11896 & 15 Nov. 2013 & [-191, 136] & 18:00            & 19:00           \\
12036 & 17 Apr. 2014 & [341, -66]  & 09:00            & 10:00           \\
12638 & 25 Feb. 2017 & [3, 388]    & 22:00            & 23:00           \\
  \noalign{\smallskip}\hline
\end{tabular}
\end{center}
\end{table}

\section{Sunspot Oscillation}
\label{sunspotosc}
\begin{figure}
\centering
\includegraphics[width=\textwidth, angle=0]{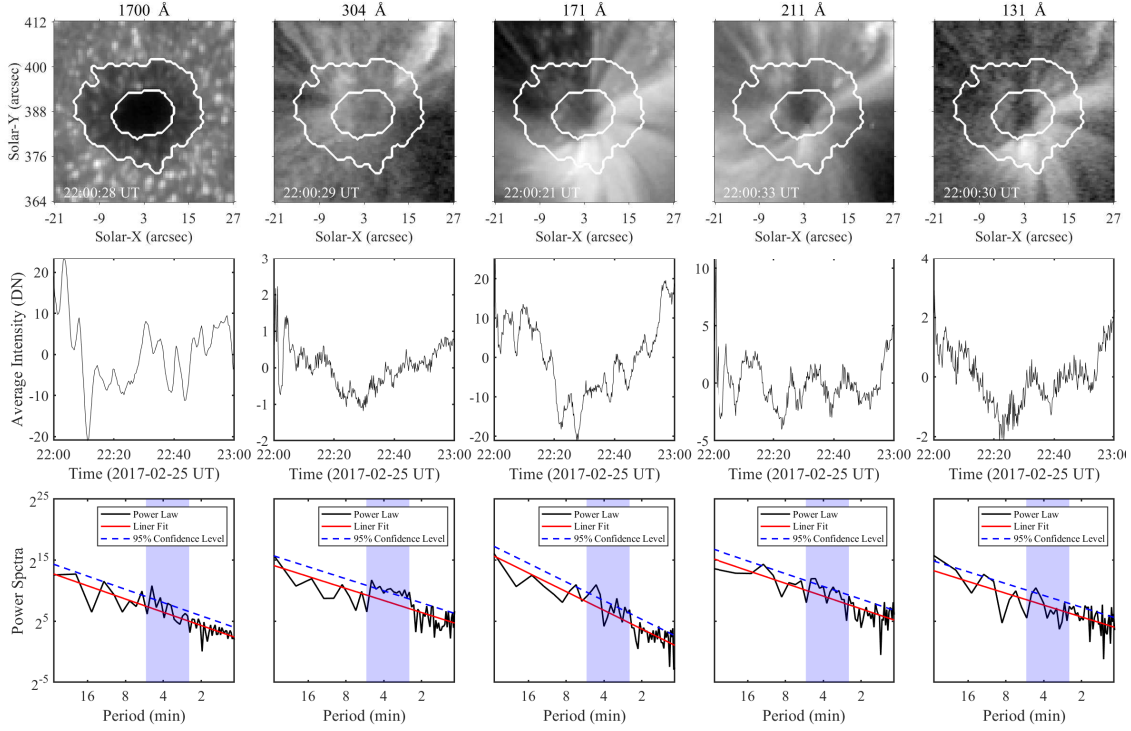}
\caption{Top:  AIA intensity images observed in NOAA 12638.
						The two white closed curves denote the umbral and penumbral boundaries, respectively.
	Middle: the average intensity curves of the entire penumbra in different channels.
	Bottom: the corresponding power spectrum of the average intensity curves, but they are a log-log scale.
					The red solid line denotes a goodness-of-fit of each power spectrum, and the blue dotted line  the 95\% confidence level.
					 Only the peaks above the confidence level are considered as the oscillation modes. The blue shadow of each spectrum denotes our region of interest whose period range from 2.5 to 5.5 minutes.}
\label{Fig1}
\end{figure}

Due to different spatial distributions of three- and five-minute oscillations in umbrae and penumbrae \citep{Balthasar+1990}, we separately studied them at different temperatures.
Using an intensity threshold to an AIA 1700 \AA\ image of different sunspots, we obtained the umbral and penumbral boundaries of each sunspot.
The top of Figure~\ref{Fig1} shows an image series observed by the AIA 1700 \AA, 304 \AA, 171 \AA, 211 \AA, and 131 \AA\ channels in NOAA 12638.
The two white closed curves denote the umbral and penumbral boundaries, respectively.
In the following sections, we take the sunspot as an example to illustrate our analysis process.

\subsection{Penumbral Oscillation}
\label{PenuOsc}
The middle row of Figure~\ref{Fig1} shows average intensity variations of the entire penumbra of the sunspot (AR 12638) on each channel.
Note that the linear trend of each curve is removed to reduce the disturbance caused by its background intensities.
The bottom of Figure~\ref{Fig1} shows the corresponding power spectra of the curves obtained by fast Fourier transform (FFT).
Note that the spectra are plotted using a log-log scale.
Moreover, we normalized the power spectra using the variance of the intensity curve shown in middle for easy comparison in different channels.
We easily find that the spectra with the log-log scale nearly exhibit a linear relationship between frequencies and powers, meaning that the powers follow a power-law distribution.

This is due to the FFT spectra of sunspot oscillations are dominated by red noise and present a power-law distribution \citep{Vaughan+2005,Ireland+etal+2015}.
For extracting true oscillation modes from the power spectra dominated by red noise, we used a similar method proposed by \citet{Vaughan+2005} to obtain goodness-of-fits of the spectra and further tested the significant peaks in the spectra.
We used linear functions to fit the log-log power spectra for obtaining their goodness-of-fits, and their 95\% confidence levels are also calculated with a chi-square test.
Red solid and blue dotted lines shown in the bottom of Figure~\ref{Fig1} indicate the goodness-of-fits for the power spectra and their confidence levels.
Here, only the peaks above the 95\% confidence level are considered as the true oscillation modes.
In our analysis, we only focus on the three- and five-minute oscillations, i.e.,  those peaks appearing between 2.5 and 5.5 minutes.
The blue regions in the bottom of Figure~\ref{Fig1}  denote the ranges between 2.5 and 5.5 minutes.
Obviously, there are several significant peaks above the confidence level in the blue shadow region of each power spectrum.
The frequency corresponding to the peaks  are 3.9 $\pm$ 0.2 mHz for AIA 1700 \AA, 3.5 $\pm$ 0.2 mHz, 5.8 $\pm$ 0.4 mHz for AIA 304 \AA, and 3.6 $\pm$ 0.2 mHz, 6.1 $\pm$ 0.1 mHz  for AIA 171 \AA. However, the peaks in AIA 211 \AA\ and 131 \AA\ are relatively weak, and their frequency ranges are 3.4 $\pm$ 0.1 and 3.6 $\pm$ 0.1 mHz, respectively.

This means that the five-minute oscillation of the penumbra distributes from AIA 1700 \AA\ to 131  \AA\ channel, indicating that the five-minute oscillation propagates from the temperature minimum to the higher corona.
But, we also find that the powers of the five-minute oscillation gradually decrease with temperature increase.
Meanwhile, the three-minute oscillation is also found in the AIA 304  \AA, 171  \AA, and 211  \AA\ channels although its powers are  weaker than the five-minute. Moreover, the powers with three-minute also appear to be a decrease trend with temperature increase.
For the sunspots in other active regions (i.e., NOAA 11176, 11433, 11479, 11896, and 12036), similar results are also found.

\subsection{Umbral Oscillation}
\label{UmbOsc}
\begin{figure}
   \centering
   \includegraphics[width=\textwidth, angle=0]{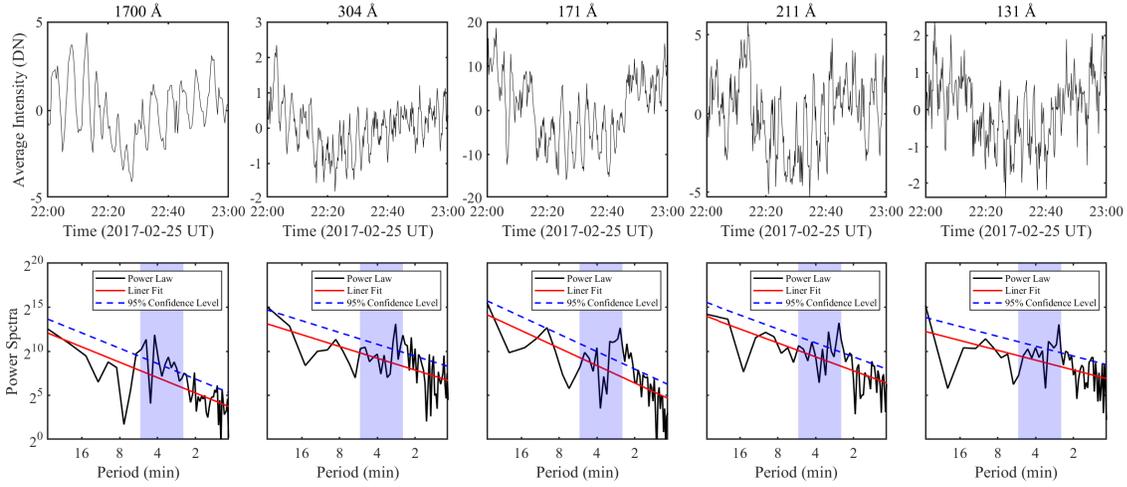}
   \caption{The average intensities of the entire umbra and their corresponding power spectra with a log-log scale.}
   \label{Fig2}
   \end{figure}

The same analysis method was used to study the oscillations in the umbra.
The top and bottom of Figure~\ref{Fig2} show the average intensity curves of the umbra and its power spectra, respectively.
To the umbra, we find that the three- and five-minute oscillations can be found, but  the three-minute oscillation exists in all AIA channels, and further their powers are relatively high, expect for AIA 1700 \AA\ channel.
However, the five-minute oscillation only exists in the AIA 1700 \AA\ channel, and the others did not.
The result indicates that, in the umbra, the three-minute oscillation distributes in different temperatures of the solar atmosphere from the temperature minimum to the higher corona.
But, the five-minute oscillation only exists in the temperature minimum.
Analyzing the sunspots of the other active regions, we also obtain similar results as above.

Combined with the oscillations observed by the six sunspots, we conclude that, from the temperature minimum to the corona,  three-minute oscillations can by observed in penumbra and umbra.  Meanwhile, we also find that the powers of three-minute oscillations are higher in umbra than in penumbrae.
To five-minute oscillations, the powers exist in penumbrae except for the temperature minimum.
Moreover, the powers of five-minute oscillations are obviously higher than that of three-minute in the temperature minimum.

\section{Spatial Distribution of Sunspot Oscillation}
\label{distribution}

To further investigate the spatial positions of three- and five-minute oscillations in sunspots, we analyzed the power spectrum of each pixel in every channel.
Here, we used two frequency bands (centered at 3.3 and 5.6 mHz with 1 mHz bandwidth) to construct their power maps.
The constructed processes are as follows:
\begin{enumerate}[(1)]
	\item Remove a linear trend of the intensity curve of each pixel  to remove the influence of the background intensity variations.
	\item Obtain the power spectrum of each intensity curve, and normalize them.
	The aim is to easily compare the power spectra to different channel data.
	\item Use a linear function to fit the power spectrum with a log-log scale, and extract the power above 95\% confidence level. The process is similar to analyzing sunspot oscillations as above.
	\item Accumulate those powers above the confidence level in a frequency band, and consider the sum as the value of the pixel position.
\end{enumerate}

\begin{figure}
   \centering
   \includegraphics[width=\textwidth, angle=0]{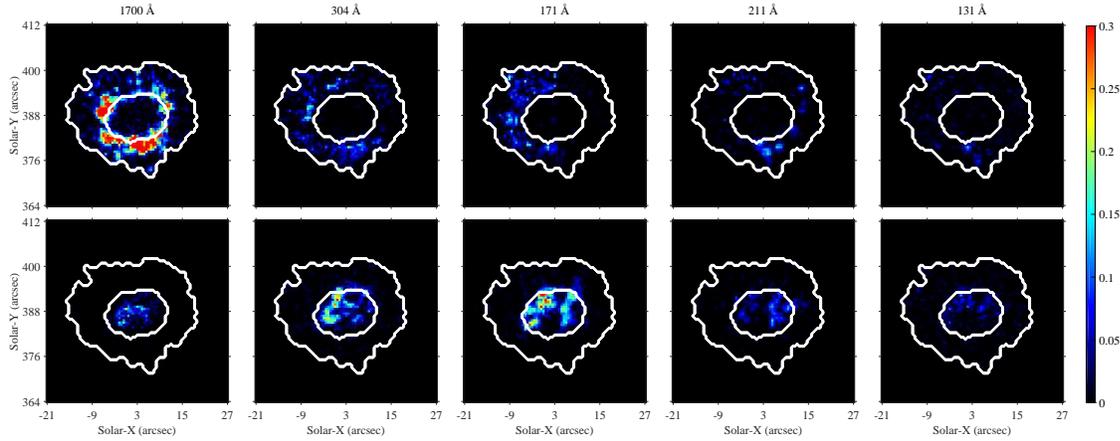}
   \caption{Top: power maps of the five-minute oscillation centered at 3.3  with 1 mHz bandwidth to the sunspot in NOAA 12638. Bottom: the map of the three-minute oscillation centered at 5.6  with 1 mHz width. The colorbar is shown on the right. The white closed curves denote the umbral and penumbral boundaries, respectively. }
   \label{Fig3}
   \end{figure}

The top and bottom of Figure~\ref{Fig3} show power maps of the sunspot in NOAA 12638 in each channel for the five- and  three-minute oscillations, respectively.
The colorbar is  shown on the right, and the color from red to black represents the power from high to low.

\subsection{Spatial Distribution of Five-minute Oscillation}
\label{distribution_5}
From the top of Figure~\ref{Fig3} we can see that the powers of the five-minute oscillation mainly located at the umbral boundary in the AIA 1700 \AA\ channel, and their distribution approximates a circle-shape structure.
Moreover, part of spectra can also be found in the penumbra and umbra.
This further explains the result, illustrated in the bottoms of Figures~\ref{Fig1} and \ref{Fig2}, that the five-minute oscillation appears in the penumbra and umbra in the AIA 1700 \AA\ channel.
In the AIA 304 \AA\ channel, the diameter of the circle enlarges, meaning that the oscillation propagates toward the penumbral boundary.
In the 1700 \AA\ and 304 \AA\ channels, the maps of the other five sunspots are similar to that in NOAA 12638.
The results have also been obtained by \cite{Kolobov+etal+2016}.
They explained that five-minute oscillations propagate along inclined magnetic field lines.
Our analysis further confirms their conclusion.
Coincidentally, we used the same one of sunspot data with them, i.e., the sunspot in NOAA 11479.
To the AIA 171\AA, 211\AA, and 131 \AA\ channels, with the temperatures increase, it still expands outwards, but the oscillation becomes more disordered and their powers distinctly decrease.
In particular, we are almost hard to find the circular shape in AIA 211 \AA\ and 131 \AA.
This indicates that the five-minute oscillation is suppressed in high-temperature corona.
We also must point out that, to the two sunspots in NOAA 11433 and 12638, in  171 \AA\ channel,  only a semicircular shape is found, and their powers are sporadic in the  AIA 211 \AA\ and 131 \AA\ channels.

\subsection{Spatial Distribution of Three-minute Oscillation}
\label{distribution_3}
The bottom of Figure~\ref{Fig3} shows the power maps of the three-minute oscillation.
The  powers mostly situate in the umbral center in the 1700 \AA\ channel.
To the AIA 304 \AA\ and 171 \AA\ channels, the powers  are obviously enhanced, and the oscillation gradually moves outward and until the umbral boundary.
But, in the AIA 211 \AA\ and 131 \AA\ channels, the oscillation diffuses to the entire sunspot, and the powers also appear relatively weak.
This indicates that the three-minute oscillation is enhanced from the transition region to the lower corona however its strength gradually decreases with the temperature increase.

\begin{figure}
   \centering
   \includegraphics[width=\textwidth, angle=0]{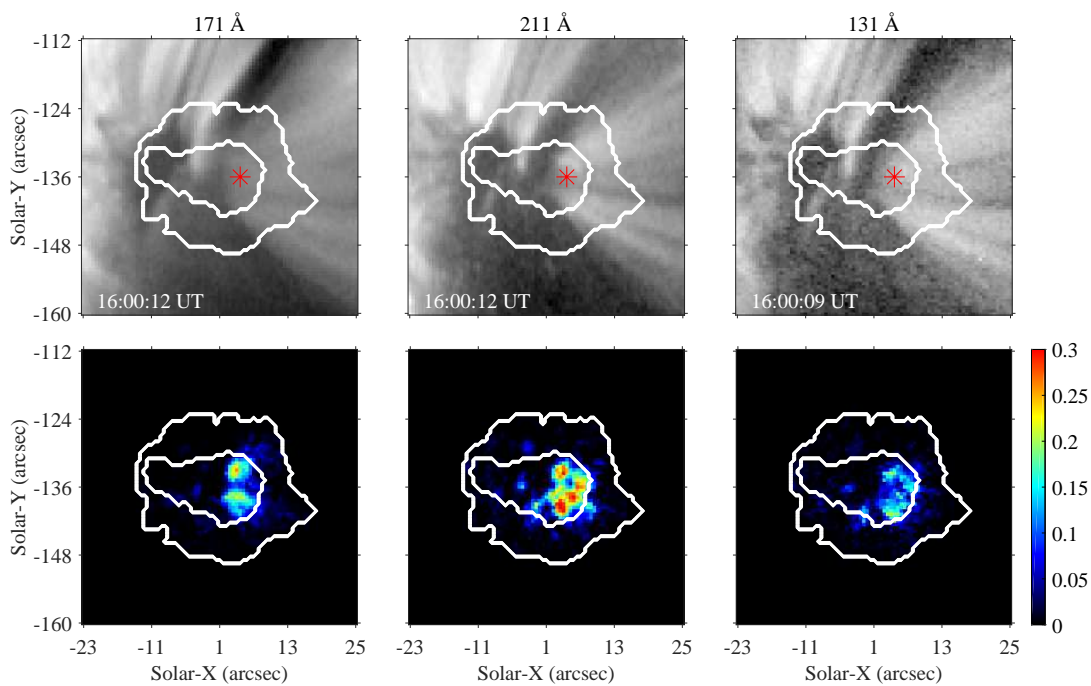}
   \caption{Top: images of the sunspot (NOAA 11176) observed in three channels: AIA 171 \AA, 211 \AA, and 131 \AA. Red asterisks denote loop footpoints. Bottom: the corresponding power maps of the three-minute oscillation. }
   \label{Fig4}
\end{figure}

\begin{figure}
   \centering
   \includegraphics[width=\textwidth, angle=0]{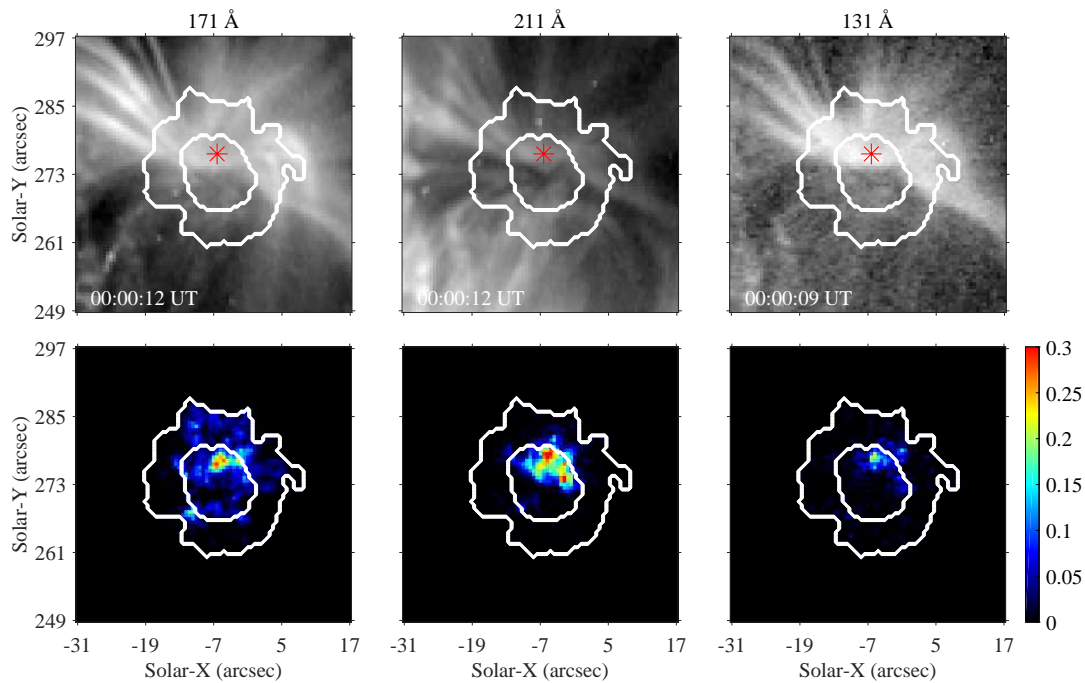}
   \caption{Same as Figure~\ref{Fig4}, but the sunspot in NOAA 11479.}
   \label{Fig5}
\end{figure}

Other sunspots also have the same results that, from the AIA 1700 \AA\ to 171  \AA\ channels, the relatively high powers gradually spread out from the umbral center to the  boundary.
Compared with the power maps of NOAA 12638 in the AIA 171 \AA, 211 \AA, and 131 \AA, the oscillations of the two sunspots in NOAA 11433 and NOAA 11479 have expended to their penumbral regions.
The original images in the two active regions and the corresponding power maps of the three-minute oscillations are shown in Figures~\ref{Fig4} and~\ref{Fig5}, respectively.
Here, the red asterisks mark the positions of the coronal loop footpoints.
Combining the intensity images and power maps, we can see that the oscillation powers outside the umbra locate at the coronal fan-structures.
Furthermore, we note that the regions of the relatively high powers roughly coincide with loop footpoints.
This meaning that three-minute oscillations may propagate along coronal fan-structures.
The conclusion is also found by some authors \citep{Reznikova+Shibasaki+2012,Yuan+Nakariakov+2012,DeMoortel+etal+2002,Jess+etal+2012}.

\section{Conclusion}
\label{conclusion}
By analyzing power spectra and power maps of three- and five-minute oscillations in randomly-chosen six sunspots, we found that the three-minute oscillation is concentrated within an umbral area in the temperature minimum, and the oscillation areas constantly increase in the transition region.
With the temperatures increase, the power spectra of the three-minute oscillation are mostly concentrated in the position of the loop footpoints, and part of the power spectra are located at fan-loop structures.
To the five-minute oscillation, the power spectra exhibit a circle-shape structure at an umbra-penumbra boundary in the temperature minimum.
From the temperature minimum to the lower corona, the oscillations gradually move outward, and disappear in the higher corona.

\begin{acknowledgements}
The authors gratefully acknowledge the anonymous referee for his/her critical reading and invaluable comments and suggestions.
We thank the use of \textit{SDO/AIA} image obtained courtesy of NASA/SDO and the AIA science teams.
This research is supported by the Joint Funds of the National Natural Science Foundation of China, and the Key Applied Basic Research Program of Yunnan Province (2018FA035), as well as the National Natural Science Foundation of China (11761141002, 11873089), the Youth Innovation Promotion Association CAS.
\end{acknowledgements}

\bibliographystyle{raa}
\bibliography{ms0172bibtex}

\begin{thebibliography}{29}
\providecommand\natexlab[1]{#1}
\providecommand\JournalTitle[1]{#1}

\bibitem[{Balthasar}(1990)]{Balthasar+1990}
{Balthasar}, H. 1990, \solphys, 125, 31

\bibitem[{Beckers} \& {Schultz}(1972)]{Beckers+Schultz+1972}
{Beckers}, J.~M., \& {Schultz}, R.~B. 1972, \solphys, 27, 61

\bibitem[{Bhatnagar} {et~al.}(1972)]{Bhatnagar+etal+1972}
{Bhatnagar}, A., {Livingston}, W.~C., \& {Harvey}, J.~W. 1972, \solphys, 27, 80

\bibitem[{Bhatnagar} \& {Tanaka}(1972)]{Bhatnagar+Tanaka+1972}
{Bhatnagar}, A., \& {Tanaka}, K. 1972, \solphys, 24, 87

\bibitem[{Bogdan} \& {Judge}(2006)]{Bogdan+Judge+2006}
{Bogdan}, T.~J., \& {Judge}, P.~G. 2006, Philosophical Transactions of the
  Royal Society of London Series A, 364, 313

\bibitem[{De Moortel} {et~al.}(2002)]{DeMoortel+etal+2002}
{De Moortel}, I., {Ireland}, J., {Hood}, A.~W., \& {Walsh}, R.~W. 2002, \aap,
  387, L13

\bibitem[{De Moortel} \& {Nakariakov}(2012)]{DeMoortel+Nakariakov+2012}
{De Moortel}, I., \& {Nakariakov}, V.~M. 2012, Philosophical Transactions of
  the Royal Society of London Series A, 370, 3193

\bibitem[{Fludra}(2001)]{Fludra+2001}
{Fludra}, A. 2001, \aap, 368, 639

\bibitem[{Giovanelli}(1972)]{Giovanelli+1972}
{Giovanelli}, R.~G. 1972, \solphys, 27, 71

\bibitem[{Ireland} {et~al.}(2015)]{Ireland+etal+2015}
{Ireland}, J., {McAteer}, R.~T.~J., \& {Inglis}, A.~R. 2015, \apj, 798, 1

\bibitem[{Jess} {et~al.}(2012)]{Jess+etal+2012}
{Jess}, D.~B., {De Moortel}, I., {Mathioudakis}, M., {et~al.} 2012, \apj, 757,
  160

\bibitem[{Jess} {et~al.}(2015)]{Jess+etal+2015}
{Jess}, D.~B., {Morton}, R.~J., {Verth}, G., {et~al.} 2015, \ssr, 190, 103

\bibitem[{Kobanov} {et~al.}(2013)]{Kobanov+etal+2013}
{Kobanov}, N.~I., {Chelpanov}, A.~A., \& {Kolobov}, D.~Y. 2013, \aap, 554, A146

\bibitem[{Kobanov} {et~al.}(2011)]{Kobanov+etal+2011}
{Kobanov}, N.~I., {Kolobov}, D.~Y., {Chupin}, S.~A., \& {Nakariakov}, V.~M.
  2011, \aap, 525, A41

\bibitem[{Kolobov} {et~al.}(2016)]{Kolobov+etal+2016}
{Kolobov}, D.~Y., {Chelpanov}, A.~A., \& {Kobanov}, N.~I. 2016, \solphys, 291,
  3339

\bibitem[{Lemen} {et~al.}(2012)]{Lemen+etal+2012}
{Lemen}, J.~R., {Title}, A.~M., {Akin}, D.~J., {et~al.} 2012, \solphys, 275, 17

\bibitem[{Lites}(1992)]{Lites+1992}
{Lites}, B.~W. 1992, in NATO Advanced Science Institutes (ASI) Series C, Vol.
  375, NATO Advanced Science Institutes (ASI) Series C, ed. J.~H. {Thomas} \&
  N.~O. {Weiss}, 261

\bibitem[{Lites} {et~al.}(1982)]{Lites+etal+1982}
{Lites}, B.~W., {White}, O.~R., \& {Packman}, D. 1982, \apj, 253, 386

\bibitem[{Marco} {et~al.}(1996)]{Marco+etal+1996}
{Marco}, E., {Aballe Villero}, M.~A., {Vazquez}, M., \& {Garcia de La Rosa},
  J.~I. 1996, \aap, 309, 284

\bibitem[{Nagashima} {et~al.}(2007)]{Nagashima+etal+2007}
{Nagashima}, K., {Sekii}, T., {Kosovichev}, A.~G., {et~al.} 2007, \pasj, 59,
  S631

\bibitem[{O'Shea} {et~al.}(2002)]{O'Shea+etal+2002}
{O'Shea}, E., {Muglach}, K., \& {Fleck}, B. 2002, \aap, 387, 642

\bibitem[{Pesnell} {et~al.}(2012)]{Pesnell+etal+2012}
{Pesnell}, W.~D., {Thompson}, B.~J., \& {Chamberlin}, P.~C. 2012, \solphys,
  275, 3

\bibitem[{Reznikova} \& {Shibasaki}(2012)]{Reznikova+Shibasaki+2012}
{Reznikova}, V.~E., \& {Shibasaki}, K. 2012, \apj, 756, 35

\bibitem[{Rouppe van der Voort} {et~al.}(2003)]{RouppevanderVoort+etal+2003}
{Rouppe van der Voort}, L.~H.~M., {Rutten}, R.~J., {S{\"u}tterlin}, P.,
  {Sloover}, P.~J., \& {Krijger}, J.~M. 2003, \aap, 403, 277

\bibitem[{Solanki}(2003)]{Solanki+2003}
{Solanki}, S.~K. 2003, \aapr, 11, 153

\bibitem[{Sych} \& {Nakariakov}(2008)]{Sych+Nakariakov+2008}
{Sych}, R.~A., \& {Nakariakov}, V.~M. 2008, \solphys, 248, 395

\bibitem[{Vaughan}(2005)]{Vaughan+2005}
{Vaughan}, S. 2005, \aap, 431, 391

\bibitem[{Yuan}(2015)]{Yuan+2015}
{Yuan}, D. 2015, Research in Astronomy and Astrophysics, 15, 1449

\bibitem[{Yuan} \& {Nakariakov}(2012)]{Yuan+Nakariakov+2012}
{Yuan}, D., \& {Nakariakov}, V.~M. 2012, \aap, 543, A9

\end{thebibliography}

\end{document}